\documentstyle[12pt,aps,psfig]{revtex}
\begin{document}
\title{Zeeman response of $d$-wave superconductors: Born
approximation for impurity and spin-orbit scattering potentials.} 
\author{C. Grimaldi\footnote{Tel: +41-21-6935825; Fax: +41-21-6933866;
E-mail: claudio.grimaldi@epfl.ch}} 
\address{\'Ecole Polytechnique F\'ed\'erale de Lausanne,
D\'epartement de microtechnique IPM,
CH-1015 Lausanne, Switzerland}

\maketitle

\centerline \\

\begin{abstract}
The effects of impurity and spin-orbit scattering potentials
can strongly affect the Zeeman response of a $d$-wave superconductor.
Here, both the phase diagram and the quasiparticle density of states 
are calculated within the Born approximation and it is found that the
spin-orbit interaction influences in a qualitatively different
way the Zeeman response of  $d$-wave
and $s$-wave superconductors.
\end{abstract}

\centerline \\


\centerline \\

{\it keywords}: $d$-wave superconductivity; spin-orbit coupling; 
Zeeman splitting.

\section{introduction}
\label{intro}

The layered structure of cuprates makes these materials
good candidates for observing Zeeman response to a magnetic field 
${\bf H}$ directed parallel to the Cu-O
planes \cite{yang,maki}. Moreover, the $d_{x^2 -y^2}$ symmetry 
of the order parameter (hereafter $d$-wave) leads in principle 
to substantial differences
with respect to the Zeeman response of isotropic $s$-wave
superconductors. For example, at zero temperature,
the tunneling conductance $\sigma_s(0)$ of a $d$-wave 
superconductor-insulator-metal junction is nonzero for finite
voltages $V$ provided ${\bf H}\neq 0$ \cite{yang,maki}, 
in sharp contrast to ordinary isotropic
$s$-wave junctions for which $\sigma_s(0)$ is zero for $V<\Delta/e$, where
$\Delta$ is the energy gap and $e$ is the electron charge \cite{fulde2}.
On the other hand, the phase diagrams of pure $s$-wave and
$d$-wave superconductors in the presence of a Zeeman magnetic field
have similar qualitative behaviours. For example, for both symmetries
of the order parameter, a first order phase transition to the normal 
state is found at low temperatures and for sufficiently strong magnetic 
fields \cite{yang,clogston}.
However there are quantitative differences. For example, at $T=0$,
the critical field is $\mu_B H_c/\Delta_0=1/\sqrt{2}$ for $s$-wave 
\cite{clogston,sarma} and
$\mu_B H_c/\Delta_0\simeq 0.56$ for $d$-wave \cite{yang,maki}, where $\Delta_0$
is the zero temperature order parameter without magnetic field.

So far, systematic theoretical studies of the Zeeman response of
anisotropic superconductors have been focused on the clean limit
of $d$-wave BCS formulation. A more realistic situation
would require the inclusion of impurity effects, since these are known
to have important effects on both thermodynamic and spectral 
quantities \cite{hirsch}.
Moreover, in addition to the disorder potential, the quasiparticles are also
spin-orbit coupled to the impurities, so that the Zeeman response
is affected by spin-mixing processes. An additional source for spin-orbit
effects could be provided by the electric fields in the vicinity of
the conducting Cu-O layers and the charge reservoirs interfaces.

Although, in the past years, the effect of spin-orbit 
coupling has been largely studied for isotropic $s$-wave superconductors
\cite{fulde2,fulde},
the corresponding situation for $d$-wave superconductors (or other
anisotropic symmetries) is still unknown. 
However, it is expected that the spin-orbit effects on the
Zeeman response of $d$-wave superconductors differ from those
of $s$-wave superconductors in a qualitative way.
In fact, already at zero magnetic field, the spin-orbit scattering is
pair-breaking and reduces both the critical temperature $T_c$ and the
order parameter \cite{grima}. 
As a consequence, for $H\neq 0$, the pair-breaking 
effects of both the external magnetic field and the spin-orbit coupling 
add together. This situation must be contrasted with the $s$-wave case,
where the spin-orbit potential is not pair-breaking and competes with 
the Zeeman response reducing the pair-breaking effect of the magnetic 
field \cite{fulde2}.

In this paper, the effects of both impurity and spin-orbit scattering
potentials are studied within a self-consistent Born approximation for
$d$-wave superconductors.
Both thermodynamic and spectral properties are investigated and
compared with those of $s$-wave superconductors.

\section{Born approximation}
\label{bornsec}

Let us consider a two-dimensional system with electrons (holes)
moving in the $x$-$y$ plane under the effect of an external
magnetic field ${\bf H}$ directed along the plane. In this
situation, the coupling of the orbital motion of the charge carriers
to the magnetic field is vanishingly small. 
In the following, no particular pairing mechanism is assumed
and the condensate will be described within the BCS formalism.
In this framework, the hamiltonian is:
\begin{equation}
\label{hami0}
H_0=\sum_{{\bf k},\alpha}\epsilon({\bf k})c^{\dagger}_{{\bf k}\alpha}
c_{{\bf k}\alpha} -I\sum_{{\bf k}\alpha}\alpha 
c^{\dagger}_{{\bf k}\alpha}c_{{\bf k}\alpha}
-\sum_{{\bf k}}\Delta({\bf k})(c^{\dagger}_{{\bf k}\uparrow}
c^{\dagger}_{-{\bf k}\downarrow}+
c_{-{\bf k}\downarrow}c_{{\bf k}\uparrow}),
\end{equation}
where $I=\mu_BH$ and $\mu_B$ is the Bohr magneton.
For a $d_{x^2-y^2}$ symmetry of the gap, $\Delta({\bf k})$
is parametrized as follows:
\begin{equation}
\label{gap1}
\Delta({\bf k})=\Delta\cos(2\phi),
\end{equation}
where $\phi$ is the polar angle in the $k_x$-$k_y$ plane.
In Eq.(\ref{hami0}), $\uparrow$ and $\downarrow$ refer
to the spin direction along and opposite to the direction
of ${\bf H}$ and it is assumed that ${\bf H}$
is directed along the $x$ direction, so that
${\bf H}=H{\bf \hat{x}}$.

The interaction hamiltonian describing the coupling to the impurities 
located randomly at ${\bf R}_i$ is given below:
\begin{eqnarray}
\label{hamiimp}
H'&=&v_{imp}\sum_{{\bf k}{\bf k}'i}\sum_{\alpha}
e^{-i({\bf k}-{\bf k}')\cdot{\bf R}_i}
c^{\dagger}_{{\bf k}\alpha}c_{{\bf k}'\alpha} \nonumber \\
&+&i \frac{v_{so}}{k_F^2}\sum_{{\bf k}{\bf k}, i}\sum_{\alpha,\beta}
e^{-i({\bf k}-{\bf k}')\cdot{\bf R}_i}
([{\bf k}\times{\bf k}']\cdot
\mbox{\boldmath $\sigma$}_{\alpha\beta})
c^{\dagger}_{{\bf k}\alpha}c_{{\bf k}'\beta},
\end{eqnarray}
where $v_{imp}$ and $v_{so}$ refer to the non-magnetic
and spin-orbit coupling to the impurities, respectively
($k_F$ is the Fermi momentum). From the Elliott relation \cite{elliott}, the
impurity and spin-orbit potentials are roughly given by
$v_{so}\sim \Delta g v_{imp}$, where $\Delta g$ is the shift
of the g-factor which, for cuprates, is of order $0.1$.
Here, however, $v_{imp}$ and $v_{so}$ will be treated as
independent variables.
Note that, since the momenta ${\bf k}$ and ${\bf k}'$
are defined in the $x$-$y$ plane,
the spin-momentum dependence of the spin-orbit
interaction simplifies to:
\begin{equation}
\label{simply}
[{\bf k}\times{\bf k}']\cdot
\mbox{\boldmath $\sigma$}_{\alpha\beta}=
[{\bf k}\times{\bf k}']\cdot {\bf \hat{z}}\,\sigma^z_{\alpha\beta}.
\end{equation}
Since the spins have been quantized along the $x$-axis,
the spin-orbit coupling leads to scattering
events always accompanied by spin-flip transitions. 

The following analysis is simplified by introducing
the usual four-components field operators \cite{fulde,rick}:
\begin{equation}
\label{four}
\Psi_{{\bf k}}=\left[\begin{array}{c}
c_{{\bf k}\uparrow} \\
c_{-{\bf k}\downarrow} \\
c^{\dagger}_{{\bf k}\uparrow} \\
c^{\dagger}_{-{\bf k}\downarrow} \end{array}
\right];\,\, 
\Psi^{\dagger}_{{\bf k}}=\left[c^{\dagger}_{{\bf k}\uparrow},
c^{\dagger}_{-{\bf k}\downarrow},c_{{\bf k}\uparrow},
c_{-{\bf k}\downarrow}\right].
\end{equation}
From Eqs.(\ref{hami0},\ref{hamiimp}) it is possible
to evaluate the equation of
motion of the field operator $\Psi_{{\bf k}}$ in the
imaginary time $\tau$:
\begin{eqnarray}
\label{motion}
\frac{d\Psi_{{\bf k}}}{d\tau}=
&-&\epsilon({\bf k})\rho_3\Psi_{{\bf k}}-
\Delta({\bf k})\rho_2\tau_2\Psi_{{\bf k}}+
I\rho_3\tau_3\Psi_{{\bf k}} \nonumber \\
&-&\sum_{{\bf k}',i}e^{i({\bf k}-{\bf k}')\cdot{\bf R}_i}
\left[v_{imp}\rho_3
+i\frac{v_{so}}{k_F^2}[{\bf k}\times{\bf k}']\cdot
{\bf \hat{z}}\,\tau_1\right]\Psi_{{\bf k}'},
\end{eqnarray}
where the products $\rho_i\tau_j$ are $4\times 4$ matrices acting
on the field operators (\ref{four}). They are constructed
by treating the Pauli matrices $\tau_j$  as elements
of the Pauli matrices $\rho_i$ as shown in the example below:
\begin{equation}
\label{example}
\rho_2\tau_2=\left(\begin{array}{cc}
0 & -i\tau_2 \\
i\tau_2 & 0 \end{array}\right).
\end{equation} 

Equation (\ref{motion}) permits to evaluate
the equation of motion of the generalized Green's function
defined as:
\begin{equation}
\label{green0}
G({\bf k},{\bf k}';\tau)=
-\langle T_{\tau} \Psi_{{\bf k}}(\tau)
\Psi^{\dagger}_{{\bf k}'}(0)\rangle,
\end{equation}
where $T_{\tau}$ is the $\tau$-order operator. 
It is straightforward to obtain from 
Eqs.(\ref{motion},\ref{green0}) the equation satisfied by
the generalized Green's function in the Matsubara frequencies 
$\omega_n=(2n+1)\pi T$:
\begin{eqnarray}
\label{green1}
G({\bf k},{\bf k}';i\omega_n)&=&
\delta_{{\bf k},{\bf k}'}G_0({\bf k};i\omega_n)+
G_0({\bf k};i\omega_n) \nonumber \\
&\times &\sum_{{\bf k}''}
\sum_ie^{i({\bf k}-{\bf k}'')\cdot{\bf R}_i}V({\bf k},{\bf k}'')
G({\bf k}'',{\bf k}';i\omega_n), \nonumber \\
\end{eqnarray}
where
\begin{equation}
\label{pot}
V({\bf k},{\bf k}'')=
v_{imp}\rho_3
+i\frac{v_{so}}{k_F^2}[{\bf k}\times{\bf k}'']\cdot
{\bf \hat{z}}\,\tau_1 ,
\end{equation}
and
\begin{equation}
\label{green2}
G_0({\bf k};i\omega_n)=[i\omega_n-\epsilon({\bf k})\rho_3-
\Delta({\bf k})\rho_2\tau_2-I\rho_3\sigma_3]^{-1} ,
\end{equation}
is the Green's function in the absence of impurities.

\begin{figure}
\protect
\centerline{\psfig{figure=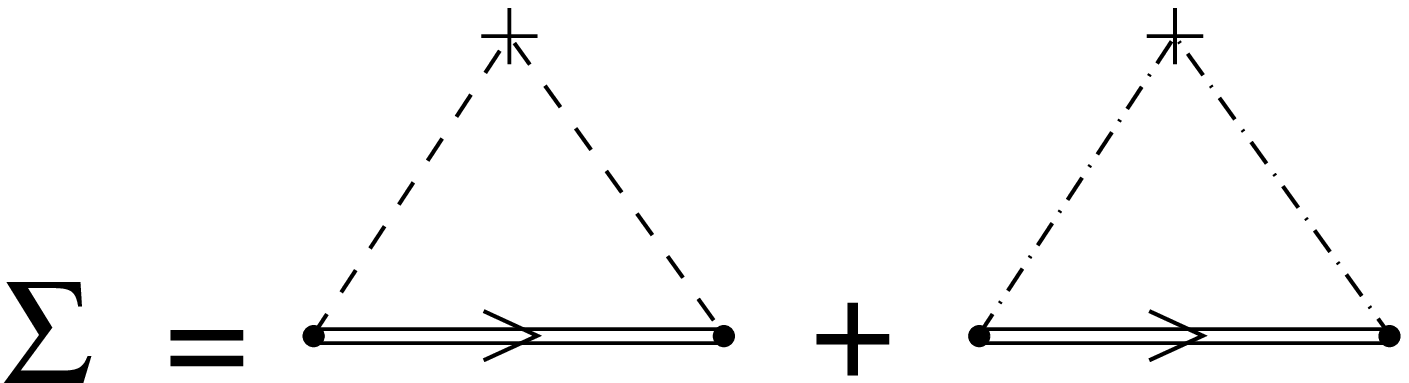,width=10cm}}
\caption{Feynmann diagrams for the self-energy in the self-consistent
Born approximation. The impurity and spin-orbit interactions are
represented by dashed and dot-dashed lines, respectively.}
\label{bornfig1}
\end{figure}

The average over all the impurity configurations of Eq.(\ref{green1})
leads to the averaged Green's function $\bar{G}$ which
satisfies the following Dyson equation \cite{rick}:
\begin{equation}
\label{green3}
\bar{G}^{-1}({\bf k},i\omega_n)=G_0^{-1}({\bf k},i\omega_n)-
\Sigma({\bf k},i\omega_n),
\end{equation}
where $\Sigma$ is the electron self-energy resulting from 
the average procedure. In this paper, $v_{imp}$ and $v_{so}$
are assumed to be sufficiently weak to justify a self-consistent
Born approximation for the self-energy $\Sigma$.
Because of
the momentum dependence of the spin-orbit interaction, the
Feynmann diagrams describing the Born approximation do not 
involve impurity--spin-orbit mixed terms and $\Sigma$ is given by
the diagrams showed in Fig. \ref{bornfig1}. Therefore,
by using Eq.(\ref{pot}), the self-consistent Born approximation
for $\Sigma$ reads:
\begin{eqnarray}
\label{self}
\Sigma({\bf k};i\omega_n)&=& n_i \sum_{{\bf k}'}
V({\bf k},{\bf k}')\bar{G}({\bf k}',i\omega_n)
V({\bf k}',{\bf k}) \nonumber \\
&=& n_iv^2_{imp}\sum_{{\bf k}'}\rho_3 
\bar{G}({\bf k}';i\omega_n)\rho_3
+n_i\frac{v^2_{so}}{k_F^4}
\sum_{{\bf k}'}|{\bf k}\times{\bf k}'|^2
\tau_1\bar{G}({\bf k}';i\omega_n)\tau_1 ,
\end{eqnarray}
where $n_i$ is the concentration of impurities. 

Equations (\ref{green3}) and (\ref{self}) must be solved
self-consistently and  the solution can be written in terms
of the following renormalized Green's function \cite{fulde,rick,maki2}:
\begin{equation}
\label{green4}
\bar{G}^{-1}({\bf k};i\omega_n)=
i(\tilde{\omega}-i\tilde{I}\rho_3\sigma_3)-
\rho_3(\tilde{\epsilon}-i\tilde{\Lambda}\rho_3\sigma_3)
-\rho_2\sigma_2(\tilde{\Delta}-i\tilde{\Omega}\rho_3\sigma_3),
\end{equation}
where for brevity the momentum and frequency
dependence of the tilded quantities has been omitted. 
The renormalized quantities can be calculated 
by substituting Eq.(\ref{green4}) into
Eqs.(\ref{green3},\ref{self}). If there is particle-hole symmetry, the 
quasiparticle dispersion remains unaffected by the presence of
impurities, {\it i.e.}, $\tilde{\epsilon}=\epsilon({\bf k})$
and $\tilde{\Lambda}=0$. For the other quantities it is useful
to introduce the variables $\tilde{\omega}_{\pm}$ and
$\tilde{\Delta}_{\pm}$ defined by:
\begin{equation}
\label{defi}
\begin{array}{ll}
\tilde{\omega}_{\pm}=\tilde{\omega}\pm i\tilde{I}\, ; & \,\,
\tilde{\Delta}_{\pm}=\tilde{\Delta}\pm i\tilde{\Omega} .
\end{array}
\end{equation}
In this way the self-consistent equations become:
\begin{equation}
\label{omega1}
\tilde{\omega}_{\pm}=
\omega_n\pm iI+n_i v_{imp}^2\sum_{{\bf k}'}
\frac{\tilde{\omega}_{\pm}}{\epsilon({\bf k}')^2+
\tilde{\Delta}_{\pm}^2+\tilde{\omega}_{\pm}^2}
+n_i\frac{v_{so}^2}{k_F^4}\sum_{{\bf k}'}
\frac{|{\bf k}\times{\bf k}'|^2\tilde{\omega}_{\mp}}{\epsilon({\bf k}')^2+
\tilde{\Delta}_{\mp}^2+\tilde{\omega}_{\mp}^2} ,
\end{equation}
\begin{equation}
\label{delta1}
\tilde{\Delta}_{\pm}=\Delta({\bf k})
+n_i v_{imp}^2\sum_{{\bf k}'}
\frac{\tilde{\Delta}_{\pm}}{\epsilon({\bf k}')^2+
\tilde{\Delta}_{\pm}^2+\tilde{\omega}_{\pm}^2}
+n_i\frac{v_{so}^2}{k_F^4}\sum_{{\bf k}'}
\frac{|{\bf k}\times{\bf k}'|^2\tilde{\Delta}_{\mp}}{\epsilon({\bf k}')^2+
\tilde{\Delta}_{\mp}^2+\tilde{\omega}_{\mp}^2}.
\end{equation}
The summations over momenta are transformed in
integrations over energy according to the usual procedure:
\begin{equation}
\label{appr}
\sum_{{\bf k}'}\rightarrow V\int\frac{d^2k'}{(2\pi)^2}
\simeq
N_0\int_0^{2\pi}\frac{d\phi'}{2\pi}\int d\epsilon ,
\end{equation}
where $N_0$ is the electronic density of states per spin state
at the Fermi level.
Performing the integration over the energy $\epsilon$,
equations (\ref{omega1}) and (\ref{delta1}) reduce to:
\begin{equation}
\label{omega2}
\tilde{\omega}_{\pm}=\omega_n\pm iI
+\frac{1}{2\tau}\int\frac{d\phi'}{2\pi}
\frac{\tilde{\omega}_{\pm}}
{[\tilde{\omega}_{\pm}^2+\tilde{\Delta}_{\pm}(\phi')^2]^{1/2}}
+\frac{1}{\tau_{so}}\int\frac{d\phi'}{2\pi}
\frac{S(\phi,\phi')\tilde{\omega}_{\mp}}
{[\tilde{\omega}_{\mp}^2+\tilde{\Delta}_{\mp}(\phi')^2]^{1/2}},
\end{equation}
\begin{equation}
\label{delta2}
\tilde{\Delta}_{\pm}(\phi)=\Delta(\phi)
+\frac{1}{2\tau}\int\frac{d\phi'}{2\pi}
\frac{\tilde{\Delta}_{\pm}(\phi')}
{[\tilde{\omega}_{\pm}^2+\tilde{\Delta}_{\pm}(\phi')^2]^{1/2}} 
+\frac{1}{\tau_{so}}\int\frac{d\phi'}{2\pi}
\frac{S(\phi,\phi')\tilde{\Delta}_{\mp}(\phi')}
{[\tilde{\omega}_{\mp}^2+\tilde{\Delta}_{\mp}(\phi')^2]^{1/2}} ,
\end{equation}
where $\tau^{-1}$ and $(\tau_{so})^{-1}$ are the scattering
rates for the non-magnetic and spin-orbit impurities, respectively.
They are given by:
\begin{equation}
\label{scatte}
\frac{1}{\tau}=2\pi n_i v_{imp}^2 N_0\, , \,\,\,
\frac{1}{\tau_{so}}=\pi n_i v_{so}^2 N_0.
\end{equation}
In Eqs.(\ref{omega2},\ref{delta2}),
the function $S(\phi,\phi')$ stems from the angular
dependence of the spin-orbit factor $|{\bf \hat{k}}\times
{\bf \hat{k}}'|^2$ by defining $\phi$ and $\phi'$ as the
polar angles of the versors ${\bf \hat{k}}$ and ${\bf \hat{k}}'$,
respectively. In explicit form, the function $S(\phi,\phi')$ is
given by:
\begin{equation}
\label{esse}
S(\phi,\phi')=\cos(\phi)^2\sin(\phi')^2+
\sin(\phi)^2\cos(\phi')^2 
-\frac{1}{2}\sin(2\phi)\sin(2\phi').
\end{equation}
The presence of such angular function leads to important differences
between non-magnetic and spin-orbit impurity effects also for
zero magnetic field.
In fact, non-magnetic 
impurities does not renormalize the gap function when this 
has $d$-wave symmetry \cite{hirsch} whereas the spin-orbit interaction,
by means of the angular function $S(\phi,\phi')$,
provides a finite renormalization. This can be readily seen
by realizing that if $\Delta(\phi)$ is of the form given
by Eq.(\ref{gap1}), then a consistent solution of Eq.(\ref{delta2})
is provided by setting $\tilde{\Delta}_{\pm}(\phi)=
\tilde{\Delta}_{\pm}\cos(2\phi)$, where $\tilde{\Delta}_{\pm}$
is the solution of:
\begin{equation}
\label{delta3}
\tilde{\Delta}_{\pm}=\Delta+
\frac{1}{\tau_{so}}\int\frac{d\phi}{2\pi}\sin(\phi)^2
\frac{\tilde{\Delta}_{\mp}\cos(2\phi)}
{[\tilde{\omega}_{\mp}^2+\tilde{\Delta}_{\mp}^2\cos(2\phi)^2]^{1/2}},
\end{equation}
and, in the same way, Eq.(\ref{omega2}) becomes:
\begin{equation}
\label{omega3}
\tilde{\omega}_{\pm}=\omega_n\pm iI
+\frac{1}{2\tau}\int\frac{d\phi}{2\pi}
\frac{\tilde{\omega}_{\pm}}
{[\tilde{\omega}_{\pm}^2+\tilde{\Delta}_{\pm}^2\cos(2\phi)^2]^{1/2}} 
+\frac{1}{\tau_{so}}\int\frac{d\phi}{2\pi}\sin(\phi)^2
\frac{\tilde{\omega}_{\mp}}
{[\tilde{\omega}_{\mp}^2+\tilde{\Delta}_{\mp}^2\cos(2\phi)^2]^{1/2}}.
\nonumber \\
\end{equation}
In obtaining Eqs.(\ref{delta3},\ref{omega3}), it has been used
the identity:
\begin{equation}
\int\frac{d\phi}{2\pi}\cos(\phi)^2 f[\cos(2\phi)]=
\int\frac{d\phi}{2\pi}\sin(\phi)^2 f[-\cos(2\phi)],
\end{equation}
where $f[\cos(2\phi)]$ is a general function of $\cos(2\phi)$.

As expected, the scalar impurity scattering contribution has
disappeared from the gap renormalization (\ref{delta3}). On the
contrary, the spin-orbit interaction modifies the gap function 
because of the presence of the angular function (\ref{esse}). 
Moreover, equations (\ref{delta3}) and (\ref{omega3}) are renormalized
in a different way by $v_{so}$ so that, even at zero magnetic field,
the spin-orbit interaction contributes to the thermodynamic and
spectral properties of $d$-wave superconductors. In fact, all the
measurable quantities can be expressed in terms of
$\tilde{u}_{\pm}=\tilde{\omega}_{\pm}/\tilde{\Delta}_{\pm}$ \cite{maki2}
which from Eqs.(\ref{delta3},\ref{omega3}) satisfies the following equation:
\begin{equation}
\label{u1}
\tilde{u}_{\pm}=\frac{\omega_n \pm iI}{\Delta}+\frac{1}{2\Delta\tau}
\int\frac{d\phi}{2\pi}
\frac{\tilde{u}_{\pm}}{[\cos(2\phi)^2+\tilde{u}_{\pm}^2]^{1/2}}
+\frac{1}{\Delta\tau_{so}}\int\frac{d\phi}{2\pi}\sin(\phi)^2
\frac{\tilde{u}_{\mp}-\tilde{u}_{\pm}\cos(2\phi)}
{[\cos(2\phi)^2+\tilde{u}_{\mp}^2]^{1/2}}.
\end{equation}
The above equation should be compared with the corresponding expression
for the two-dimensional isotropic $s$-wave case which, by setting 
$\Delta(\phi)=\Delta$
in Eqs.(\ref{omega2},\ref{delta2}), is found to be \cite{fulde2,fulde,maki2}:
\begin{equation}
\label{swave}
\tilde{u}_{\pm}=\frac{\omega_n\pm iI}{\Delta}+\frac{1}{2\Delta\tau_{so}}
\frac{\tilde{u}_{\mp}-\tilde{u}_{\pm}}{[1+\tilde{u}_{\mp}^2]^{1/2}},
\end{equation}
where the contribution of the impurity scattering has vanished
because of Anderson's theorem. When $H=0$, equation (\ref{swave})
reduces to $\tilde{u}_+=\tilde{u}_-=\omega_n/\Delta$ and does not 
depend on the spin-orbit scattering rate while Eq.(\ref{u1}) still 
depends on $\tau$ and $\tau_{so}$.
In fact, in a $d$-wave superconductor, both the non-magnetic
impurity and the spin-orbit scatterings are pair breaking and they tend
to suppress superconductivity \cite{grima}.
When ${\bf H}\neq 0$, it is therefore expected that
the Zeeman response of a $d$-wave superconductor differs qualitatively
from that of a $s$-wave condensate. 

\section{Phase diagram}
\label{phase}

Equation (\ref{u1}) permits to obtain all the informations
needed to calculate the phase diagram of a dirt $d$-wave
superconductor in a Zeeman magnetic field. Let us start by
considering the self-consistent equation for the order parameter
$\Delta$: 
\begin{equation}
\label{gap2}
\Delta=\frac{V_0}{4} T\sum_n \sum_{{\bf k}}\cos(2\phi)
\mbox{Tr}\left[\rho_2\tau_2 \bar{G}({\bf k},i\omega_n)\right]=
\lambda \pi T\sum_n \int\frac{d\phi}{2\pi}
\mbox{Re}\frac{\cos(2\phi)^2}
{[\cos(2\phi)^2+\tilde{u}_+^2]^{1/2}},
\end{equation}
where $V_0$ is the pairing interaction and $\lambda=V_0N_0$.
The summation over the frequencies is implicitly assumed to be
restricted by a cut-off energy. However, both the cut-off frequency and
the pairing interaction can be absorbed in the definition of the
critical temperature $T_{c0}$ for a pure superconductor
($\tau^{-1}=0$, $\tau_{so}^{-1}=0$) without external magnetic field.
In this way the gap equation can be rewritten as:
\begin{equation}
\label{gap3}
\ln\left(\frac{T}{T_{c0}}\right)=4\pi T\sum_{n\ge 0}
\left\{\int\frac{d\phi}{2\pi}\mbox{Re}\frac{1}{\Delta}
\frac{\cos(2\phi)^2}{[\cos(2\phi)^2+\tilde{u}_+^2]^{1/2}},
-\frac{1}{2\omega_n}\right\}
\end{equation}
In the hypothesis that the transition to the normal state is of the
second order (see below), the critical temperature $T_c$ is obtained
from Eq.(\ref{gap2}) by setting $\Delta\rightarrow 0$ and it is
given by:
\begin{equation}
\label{Tc1}
\ln\left(\frac{T}{T_{c0}}\right)=
\psi\left(\frac{1}{2}\right)-\frac{1}{2}\left[\left(1+
\frac{1}{4\tau_{so}b}\right)\psi\left(\frac{1}{2}+a+\frac{b}
{2\pi T_c}\right)+\left(1-
\frac{1}{4\tau_{so}b}\right)\psi\left(\frac{1}{2}+a-\frac{b}
{2\pi T_c}\right)\right],
\end{equation}
where $a=(\tau^{-1}+\tau_{so}^{-1})/4\pi T_c$ and
$b=[1/(4\tau_{so})^2-I^2]^{1/2}$ and $\psi$ is the di-gamma function.
When $I=\mu_BH=0$ equation (\ref{Tc1}) reduces to:
\begin{equation}
\label{Tc2}
\ln\left(\frac{T}{T_{c0}}\right)=\psi\left(\frac{1}{2}\right)-
\psi\left(\frac{1}{2}+\frac{1}{4\pi T_c\tau}
+\frac{3}{4}\frac{1}{2\pi T_c\tau_{so}}\right),
\end{equation}
which coincides with the result obtained in Ref.\cite{grima} in the weak
scattering limit \cite{note}. Equation (\ref{Tc2}) shows
that, even at zero magnetic field, the spin-orbit scattering contributes
together with the non-magnetic impurity scattering to the suppression
of $T_c$. 

\begin{figure}
\protect
\centerline{\psfig{figure=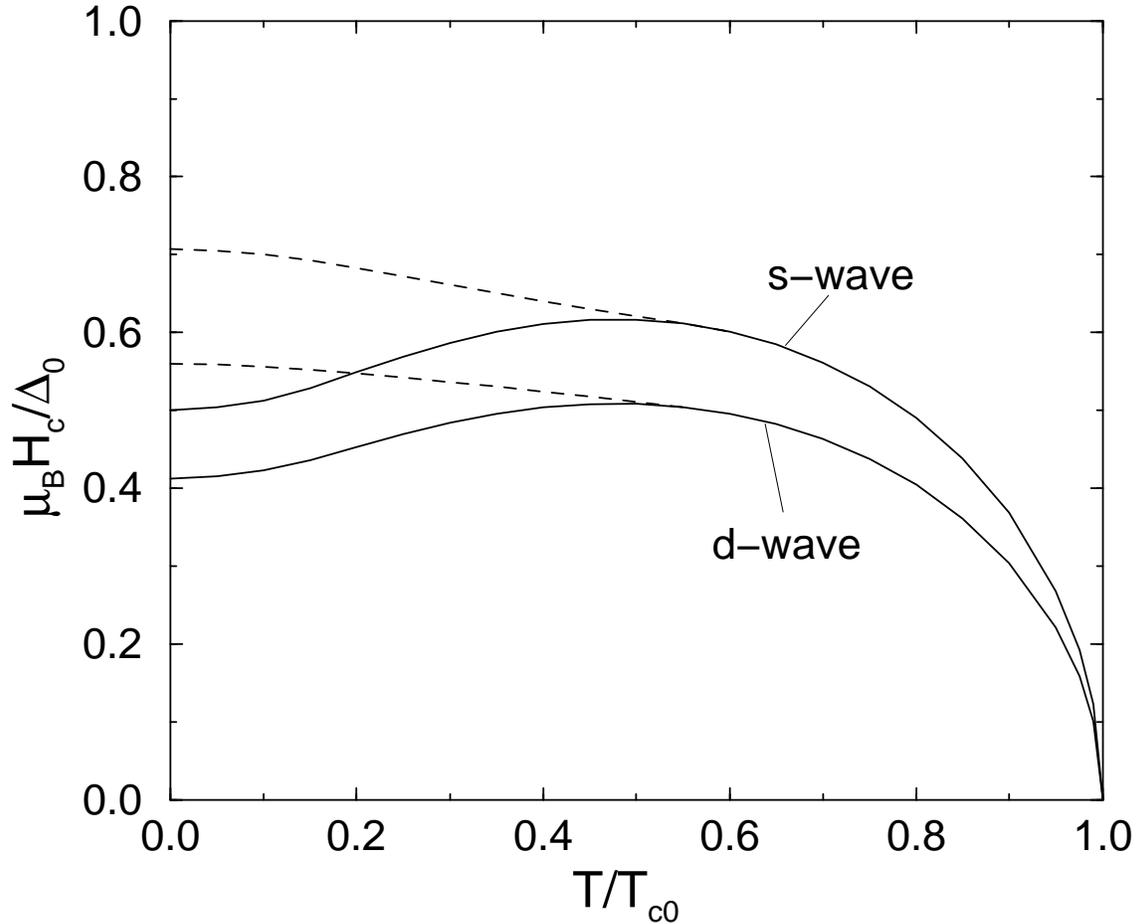,width=15cm}}
\caption{Phase diagram for pure $s$-wave and $d$-wave
superconductors in the presence of a Zeeman magnetic field.
$\Delta_0$ and $T_{c0}$ are the order parameter and the critical
temperature without the external magnetic field, respectively.
For $T/T_{c0}>0.56$ the solid lines are the second-order phase
boundary between the normal (above the solid lines) and the
superconducting (below the solid lines) states. For $T/T_{c0}<0.56$
both the $s$-wave and the $d$-wave states show a first-order
transition to the normal state marked by the dashed lines.
In this region, the solid lines represent the supercooling fields.}
\label{bornfig2}
\end{figure}

For large enough values of the external magnetic field,
the transition to the normal state becomes of first order 
\cite{clogston}.
This situation is studied by evaluating the difference of the free
energy between the superconducting and the normal states 
$\Delta F=F_s-F_n$. If, by rising the temperature and/or the 
magnetic field, $\Delta F$ changes sign while $\Delta$ remains finite,
then the system undergoes a first order phase transition to the normal
state with critical field $B_c$ and $T_c$ determined by 
$\Delta F=0$ \cite{fulde2}.
By following Ref.\cite{maki2}, $\Delta F$ is obtained by:
\begin{equation}
\label{free1}
\Delta F=\int_0^{V_0}dV_0\, \Delta^2,
\end{equation}  
and by using Eqs.(\ref{u1},\ref{gap2}) one readily finds:
\begin{equation}
\label{free2}
\Delta F=-N_0\Delta 2\pi T\sum_{n\ge 0}\int\frac{d\phi}{2\pi}\mbox{Re}
\left\{2[\cos(2\phi)^2+u_+^2]^{1/2}-2u_+-\frac{\cos(2\phi)^2}
{[\cos(2\phi)^2+u_+^2]^{1/2}}\right\}.
\end{equation}
The numerical solution of equations (\ref{Tc1}) and (\ref{free2})
are shown in Fig. \ref{bornfig2} for the pure limit 
and, for comparison, the $d$-wave
solution is plotted together with the $s$-wave one. In the phase
diagram, the solid and dashed lines are solutions 
of equations (\ref{Tc1}) and (\ref{free2}), respectively.
For both $d$-wave and $s$-wave, the transition to the normal
state is of second order for $T/T_{c0}\ge 0.56$ \cite{yang,maki,sarma}
while for lower
temperatures the transition, marked by dashed lines, becomes of first order.
For $T/T_{c0}< 0.56$ the solid lines represent the supercooling 
field \cite{yang,maki,fulde2}.
As already said in the introduction, at zero temperature the first
order transition to the normal state is obtained by critical fields
$\mu_B H_c/\Delta_0=1/\sqrt{2}$ \cite{clogston} for $s$-wave 
and $\mu_B H_c/\Delta_0\simeq 0.56$ for $d$-wave \cite{yang,maki}.
In this paper, the Fulde-Ferrel--Larkin-Ovchinnikov state \cite{fflo}
which appears at low temperatures has not been considered since 
disorder tends to restore the zero momentum pairing \cite{grunberg}.
For the pure $d$-wave case, the reader can find the phase diagram
including the non-zero momentum pairing state in Ref.\cite{yang}.

\begin{figure}
\protect
\centerline{\psfig{figure=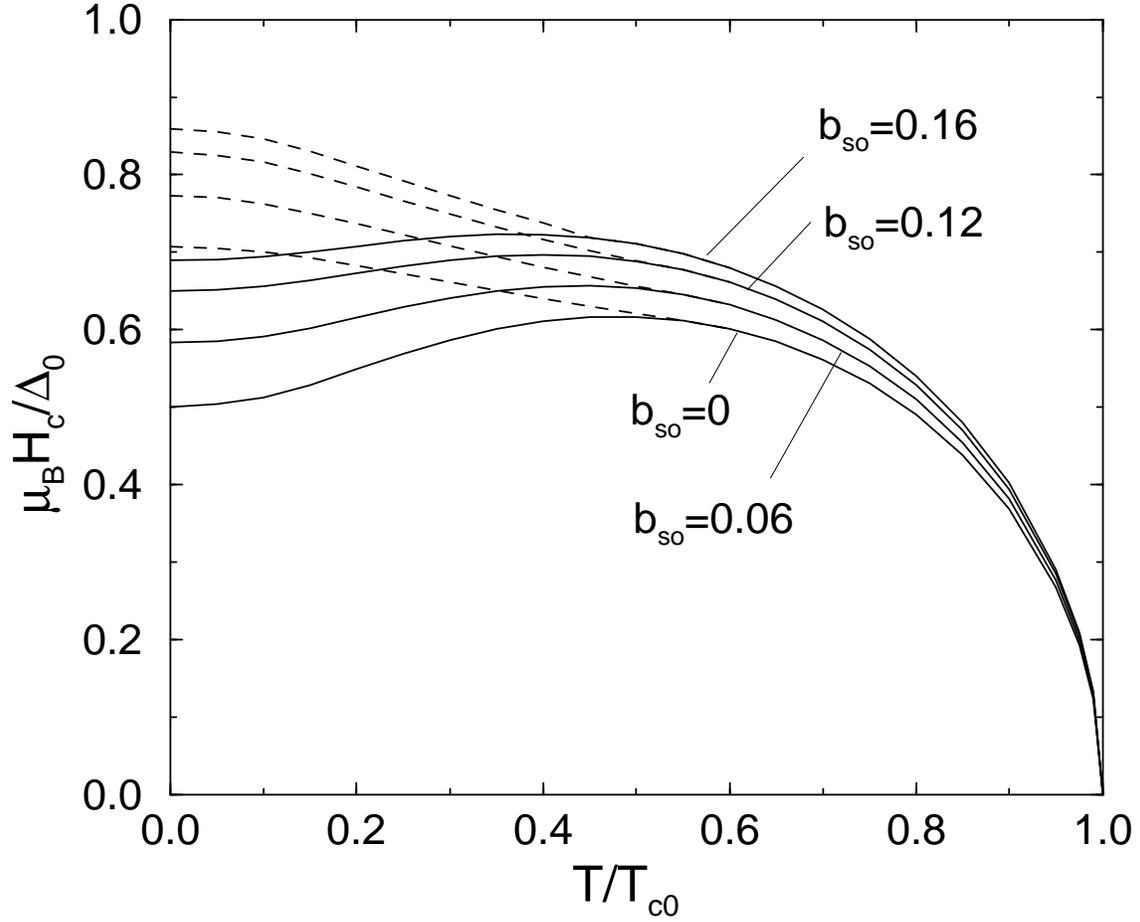,width=15cm}}
\caption{Phase diagram for a $s$-wave superconductor with
impurity and spin-orbit scattering centres. The critical field
is unaffected by the impurity potential while it increases by increasing
the spin-orbit scattering parameter $b_{so}=1/(2\Delta_0\tau_{so})$ where
$\Delta_0$ is the order parameter in the pure limit without magnetic field.
The solid and dashed lines have the same meaning as in 
Fig.\protect\ref{bornfig2}.}
\label{bornfig3}
\end{figure}

\begin{figure}
\protect
\centerline{\psfig{figure=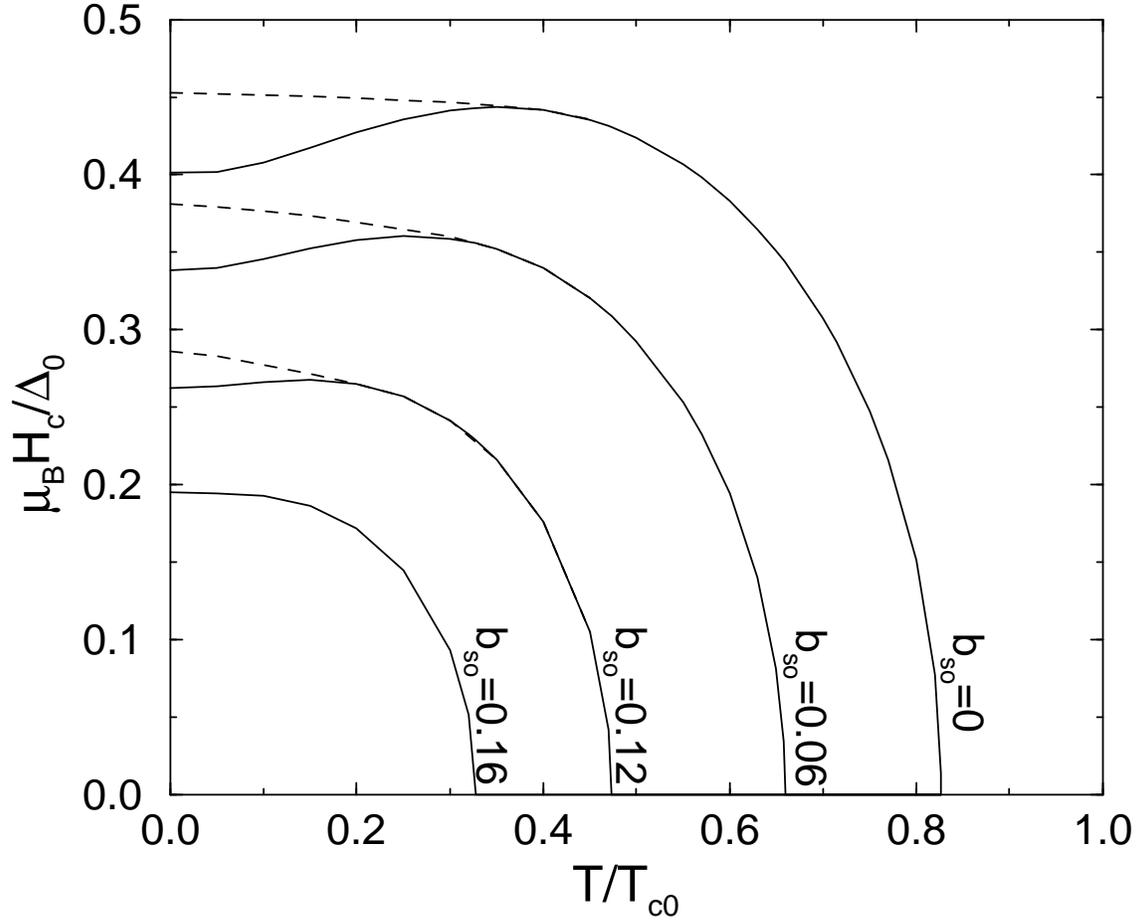,width=15cm}}
\caption{Phase diagram for a $d$-wave superconductor with
impurity and spin-orbit scattering centres. The impurity
scattering parameter is $b_n=1/(2\Delta_0\tau)=0.1$ where
$\Delta_0$ is the order parameter in the pure limit without 
magnetic field. The solid and dashed lines have the same meaning as in 
Fig.\protect\ref{bornfig2}. Note that, contrary to the case shown in
Fig.\protect\ref{bornfig3}, already for $b_{so}=0.16$ the transition
to the normal state is of second-order for all the temperature range.}
\label{bornfig4}
\end{figure}

Although for the pure limit the phase diagrams of the Zeeman
response of $s$-wave and $d$-wave superconductors are qualitatively similar,
they drastically differ when the coupling to the non-magnetic and
spin-orbit impurity scatterings is switched on. In Figs. \ref{bornfig3} 
and \ref{bornfig4},
the phase diagrams for $s$-wave and $d$-wave superconductors are plotted
for finite values of $\tau^{-1}$ and $\tau_{s0}^{-1}$. In both figures,
the impurity scattering parameter $b_n=1/(2\Delta_0\tau)$ is set equal 
to $0.1$, while the spin-orbit scattering parameter 
$b_{so}=1/(2\Delta_0\tau_{so})$ assumes four different values: $b_{so}=0,\,
0.06,\, 0.12,\, 0.16$. In the $s$-wave case, Fig. 3, the phase diagram
is insensitive to $b_n\neq 0$, while finite values of $b_{so}$ enhance
the critical field for all temperatures. Moreover, the temperature interval
of first-order phase transition (dashed lines) decreases as $b_{so}$
increases and for $b_{so}>2.32$ the transition becomes continuous for
all temperatures \cite{fulde2}. 
This behavior is due to the spin-mixing effect of
the spin-orbit interaction which lowers the Zeeman response and
consequently the depairing effect of the magnetic field.
On the other hand, in the $d$-wave case shown in Fig. 4, the spin-orbit
scattering is pair breaking and for $b_{so}>0$
the critical field is lowered. This situation can be understood by realizing
that finite values of $b_{so}$ lead to a weakening of the superconducting
state \cite{grima} so that, with respect to the $b_{so}=0$ case, 
lower values of $H$ 
are needed to suppress completely superconductivity. Another striking
feature is that, due to the nodes of the $d$-wave order parameter, the
$b_{so}$ dependence of the phase diagram is much stronger than for the $s$-wave
case. In fact, already for $b_{so}=0.16$ there is not signature for
first-order transition whereas for an $s$-wave superconductor the first-order
transition disappears only for $b_{so}>2.36$, {\it i. e.}, a difference of
one order of magnitude. 
 
It is important to stress that the remarkable difference between the
$d$-wave and $s$-wave phase diagrams has been obtained in the Born 
limit of non-magnetic and spin-orbit impurity scatterings. 
However it is well known that in high-$T_c$ superconductors the effect
of disorder is best described by the strongly resonant limit of the
impurity potential so that the Born approximation may result inadequate.
In practice, one should formulate the Zeeman response by employing
the t-matrix approximation for both the non-magnetic and the spin-orbit
potentials. Such a calculation has already been reported in Ref.\cite{grima}
for zero external magnetic field. The generalization for $H\neq 0$ is
currently under investigation and the results reported here may provide
a useful tool for testing the more general t-matrix solution.

\section{Density of states}
\label{density}

\begin{figure}
\protect
\centerline{\psfig{figure=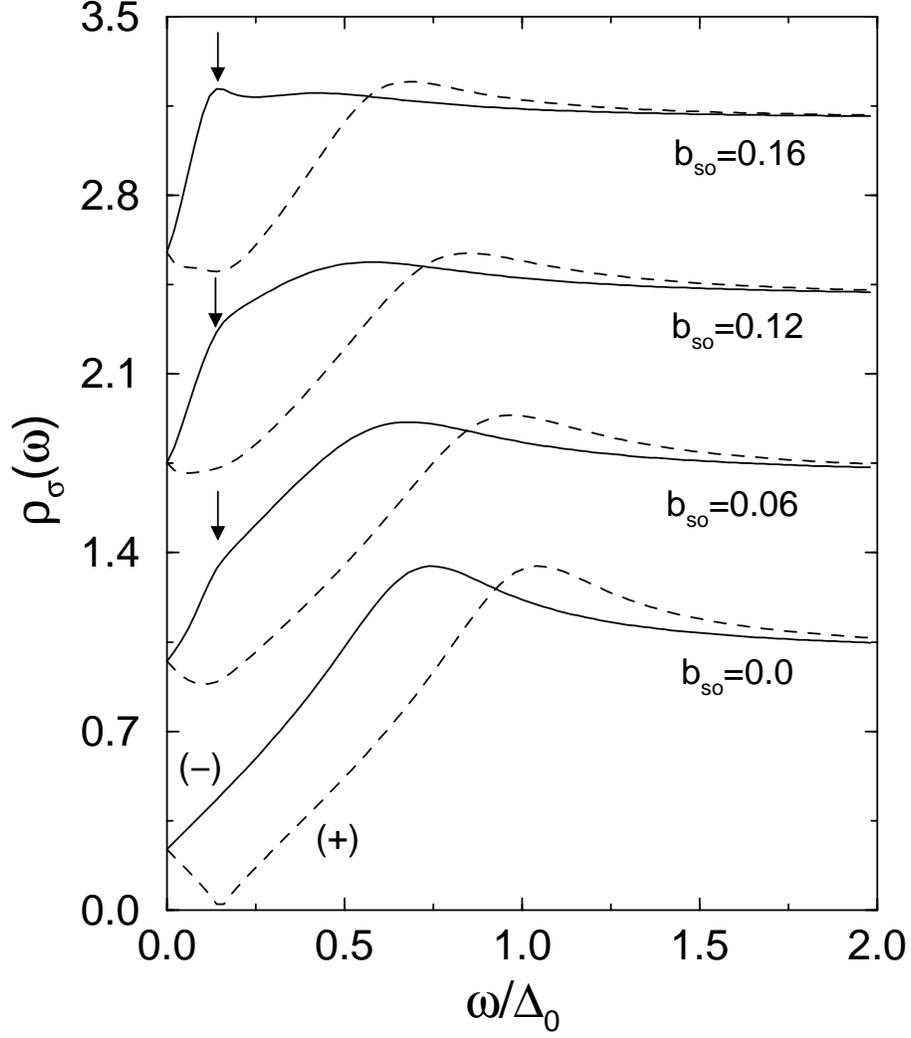,width=12cm}}
\caption{Zeeman splitted quasiparticle density of states 
$\rho_+(\omega)$ (dashed lines) and $\rho_-(\omega)$ (solid lines) for 
a $d$-wave superconductor with $b_n=0.1$, $\mu_BH/\Delta_0=0.15$
and different values of the spin-orbit scattering parameter $b_{so}$.
The curves for different values of $b_{so}$ are vertically shifted
by multiples of $0.7$. 
Note the structure (marked by the arrows) at $\omega\simeq \mu_B H$ 
which develops as $b_{so}$ increases}
\label{bornfig5}
\end{figure}

\begin{figure}
\protect
\centerline{\psfig{figure=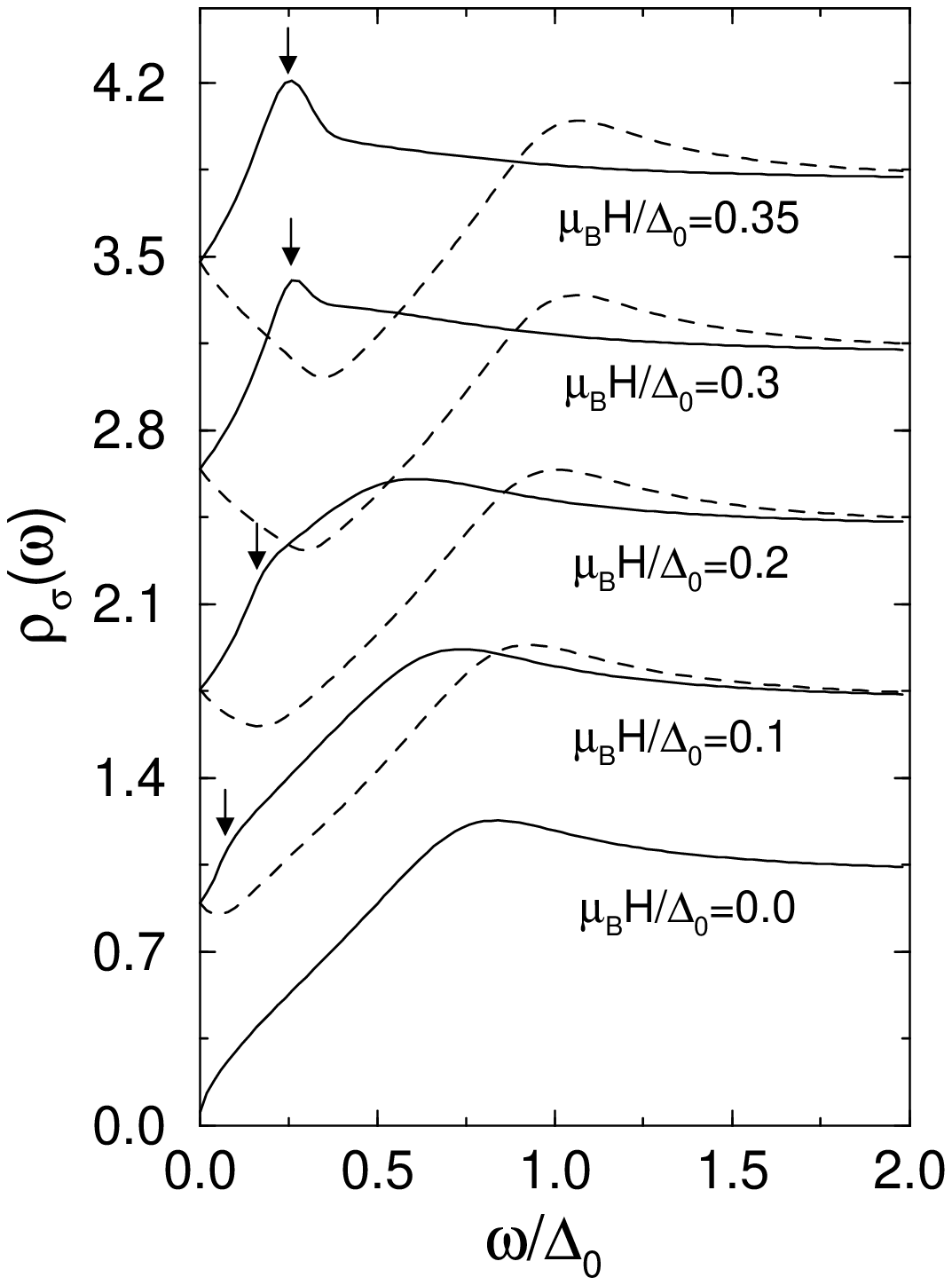,width=12cm}}
\caption{Zeeman splitted quasiparticle density of states 
$\rho_+(\omega)$ (dashed lines) and $\rho_-(\omega)$ (solid lines) for 
a $d$-wave superconductor with $b_n=0.1$, $b_{so}=0.06$
and different values of the external magnetic field.
The curves for different values of $H$ are vertically shifted
by multiples of $0.7$. 
The arrows indicate the resonant structure which develops a
peak for $\mu_BH>0.3$ (see text).}
\label{bornfig6}
\end{figure}

Having described the effect of $b_n$ and $b_{so}$ on the phase
diagram, it is interesting to investigate also how the spectral properties
are modified. To this end, equation (\ref{u1}) must be analytically
continued to the real axis by setting $i\tilde{u}_{\pm}\rightarrow u_{\pm}$
and $i\omega_n\rightarrow \omega$. In this way, the quasiparticle
density of states (DOS) per spin direction in units of the normal
state DOS $N_0$ can be calculated by the following expression:
\begin{equation}
\label{dos1}
\rho_{\pm}(\omega)=\frac{N_{\pm}(\omega)}{N_0}=
\mbox{sgn}(\omega)\int\frac{d\phi}{2\pi}
\mbox{Re}\frac{u_{\pm}}{[u_{\pm}^2-\cos(2\phi)^2]^{1/2}}.
\end{equation}
In Fig. \ref{bornfig5} it is reported the quasiparticle DOS 
for $I=\mu_BH=0.15\Delta_0$,
$b_n=0.1$ and different values of the spin-orbit parameter $b_{so}$.
For clarity, the curves with $b_{so}\neq 0$ have been vertically shifted by 
$0.7$, $2\times 0.7$ and $3\times 0.7$ with respect to those with
$b_{so}=0$. For $b_{so}=0$, the two DOS per spin state,
$\rho_+$ (dashed lines) and $\rho_-$ (solid lines), show a clear Zeeman 
splitting and for $\omega=0$
the total DOS $\rho=\rho_++\rho_-$ is different from zero as expected for
a $d$-wave superconductor. For $b_{so}>0$ the total DOS at $\omega=0$ is
enhanced at the expenses of the coherence peaks which show a decrease of
spectral weight. Moreover, at
$\omega\simeq \mu_BH$, $\rho_-$ develops a structure (marked by the arrows)
 which becomes a peak
at $b_{so}=0.16$. Such a structure is even more visible in Fig. \ref{bornfig6}
 where
$\rho_{\pm}$ is plotted for $b_n=0.1$, $b_{so}=0.06$ and different values
of the external magnetic field. 
The origin of this peak can be understood by the following reasoning.
At the Fermi wave-vector, and for a pure superconductor, the quasiparticle
energies for spin up and down are $E_{\pm}(\phi)=
\Delta|\cos(2\phi)|\pm \mu_BH$ and therefore, depending on the values of
$H$ and $\Delta$, two quasiparticles with different spin orientation and
angles $\phi$ can have equal energies.  
For example, for $\phi_1=0$ and
$\phi_2=\pi/4$, the two energies $E_-(\phi_1)$ and $E_+(\phi_2)$
are equal to $\omega=\mu_BH$ if $\Delta=2\mu_BH$.
Since the spin-orbit potential connects quasiparticle states with
different spin orientation but equal energies, the two states
$E_-(\phi_1)$ and $E_+(\phi_2)$ are coupled by the spin-orbit interaction
and a enhanced signal should be expected at $\omega\simeq \mu_BH$.
Note in fact that in Fig. \ref{bornfig6} the low energy peak is 
more pronounced for
$\mu_BH=0.35 \Delta_0$ where, since $\Delta\simeq 0.68\Delta_0$, the
condition $\Delta=2\mu_BH$ is nearly fulfilled.

\section{conclusions}
\label{conclusions}

In conclusion, it has been shown within the Born approximation that the
presence of impurity and spin-orbit scattering centres strongly affects the
Zeeman response of a $d$-wave superconductor. Both the phase diagram and
the quasiparticle density of states show features qualitatively different
from those of a $s$-wave superconductor. In fact, by increasing the value
of the spin-orbit scattering parameter $b_{so}=1/(2\Delta_0\tau_{so})$
the critical field $H_c$ is strongly lowered whereas in a $s$-wave 
superconductor $H_c$ increases. Moreover, the influence of $b_{so}$ on
the superconducting state is much stronger for the $d$-wave symmetry.
Concerning the spectral properties, the Zeeman splitted density of states
of a $d$-wave superconductor shows interesting features which are missing
in a $s$-wave superconductor. In fact, for sufficiently large values of
$b_{so}$ and/or $H$ a resonant peak develops at energies close to
$\mu_B H$. The origin of this feature is given by the anisotropy of the
order parameter and the spin-flip transitions due to the spin-orbit
scattering. An important open question concerns with the possibility of
going beyond the Born approximation and employing a t-matrix approach
for the Zeeman response in the presence of impurity and spin-orbit
scattering centres in order to test the solidity of the results here presented.

\acknowledgements
The author thanks P. Fulde and S. Str\"assler for interesting discussions.

\newpage

\newpage

\newpage

\newpage

\newpage

\begin{references}
\bibitem{note}
In Ref.\cite{grima} a different notation has been used
in which the impurity potential is parametrized by
$\Gamma=n_i/(\pi N_0)$ and $c=1/(\pi N_0v_{imp})$ and
the spin-orbit interaction is parametrized as $v_{so}=
\Delta g v_{imp}$, where $\Delta g$ is the shift of the 
electronic $g$-factor. Therefore,
by using Eq.(\ref{scatte}), $1/(2\tau)=\Gamma/c^2$ and
$1/\tau_{so}=\Gamma(\Delta g/c)^2$.
\end{references}
\end{document}